\documentclass{article}

\usepackage{arxiv}

\usepackage[utf8]{inputenc} 
\usepackage[T1]{fontenc}    
\usepackage{url}            
\usepackage{booktabs}       
\usepackage{amsfonts}       
\usepackage{nicefrac}       
\usepackage{microtype}      
\usepackage[noadjust]{cite}

\usepackage{graphicx}
\usepackage{bm}
\usepackage{amsfonts}
\usepackage{color}

\usepackage{amsmath}    
\usepackage{epsfig}
\usepackage{subfigure}  
\usepackage{hyperref}   
\usepackage{bm}
\usepackage{amssymb}

\newcommand{\be}{\begin{equation}}
\newcommand{\ee}{\end{equation}}

\newcommand{\bu}{\boldsymbol{u}}

\newcommand{\bhu}{\boldsymbol{\hat{u}}}

\newcommand{\bk}{\boldsymbol{k}}
\newcommand{\bq}{\boldsymbol{q}}
\newcommand{\bp}{\boldsymbol{p}}

\title{TURB-Rot. A large database of 3d and 2d snapshots from turbulent rotating flows}

\author{
  L. Biferale \\
  Dept. Physics and INFN\\
  University of Rome Tor Vergata, Italy, and IIC-Paris, France\\
  \texttt{biferale@roma2.infn.it} \\
   \And
  F. Bonaccorso \\
  Center for Life Nano Science@La Sapienza\\
  Istituto Italiano di Tecnologia and INFN\\
  University of Rome Tor Vergata, Italy.\\
  \texttt{fabio.bonaccorso@roma2.infn.it} \\
   \And
  M. Buzzicotti \\
  Dept. Physics and INFN\\
  University of Rome Tor Vergata, Italy.\\
  \texttt{michele.buzzicotti@roma2.infn.it} \\
   \And
  P. Clark Di Leoni \\
  Department of Mechanical Engineering,\\
  Johns Hopkins University, Baltimore, USA.\\
  \texttt{pato@jhu.edu} \\
}

\begin{document}
\maketitle

\begin{abstract}
We present TURB-Rot, a new open database of 3d and 2d snapshots of turbulent velocity fields,  obtained by  Direct Numerical Simulations (DNS) of the original Navier-Stokes equations in the presence of rotation.
The aim is to provide the community interested in data-assimilation and/or computer vision  with a  new testing-ground made of roughly 300K complex images and fields.
TURB-Rot data are characterized by multi-scales strongly non-Gaussian features and rough, non-differentiable, fields over almost two decades of scales.  In addition, coming from  fully resolved numerical simulations of the original partial differential equations, they offer the possibility to apply a wide range of approaches, from  equation-free to physics-based models. TURB-Rot data are reachable at  \url{http://smart-turb.roma2.infn.it}.

\end{abstract}


\section{Introduction}
Turbulent flows under rotation are a physical system extremely relevant in many areas of research dealing with both natural, e.g. astrophysical and geophysical \cite{barnes2001,cho2008atmospheric}, and industrial flows \cite{dumitrescu2004}.
Rotating turbulence, is well known  for its complex dynamics resulting from the fact that the energy injected from the external forcing produces  large-scale cyclonic and/or anti-cyclonic structures as well as small-scale intermittent homogeneous and quasi isotropic highly non-Gaussian fluctuations, leading to the formation of structures over many decades of scales (see \cite{alexakis2018cascades} for a recent review  and \cite{greenspan1968,davidson2013} for textbooks). In this short note, we present an open access data base of rotating turbulence. We deploy a first dataset at medium numerical resolution, planning to extend it to larger resolution and different datatype, including Lagrangian information along particle trajectories in a later stage. The aim is to have a huge set of 2d and 3d images/fields and trajectories that can be used by people belonging to  different communities and with different goals, from  data-assimilation and field reconstruction, interesting for numerical weather prediction, to denoising and inpainting  for computer vision tasks \cite{pathak2016context,yeh2017semantic,ulyanov2018deep,bertalmio2000image,osher2005iterative,little2019statistical}. A first application has been presented by us in \cite{buzzicotti2020}.

In Sec. 2 we provide a brief description of the Direct Numerical Simulations (DNS) performed to generated our TURB-Rot database while in Sec. 3 we discuss how the 300K images have been selected and how it is possible to access to  the entire data-set.

\section{Numerical Simulations}
To generate the first dataset deployed on TURB-Rot database, we have performed  a DNS of the Navier-Stokes equations (NSE) for  incompressible rotating fluid in a triply periodic domain of size $L=2\pi$, using a fully dealiased parallel 3d pseudospectral code on a grid of  $N= 256^3$ collocation points. The time integration has been implemented with a  second-order Adams-Bashforth scheme with viscous term integrated implicitly. 
The NSE for a fluid in the rotating frame can be written as
\begin{equation}
\label{eq:nse}
\begin{cases}
\partial_t \bm{u} + \bm{u} \cdot \nabla \bm{u} +2{\bm \Omega} \times {\bm u}
= - \nabla p + \nu\Delta\bm{u}+ \alpha \Delta^{-1}\bm{u} + \bm{f} \\
\nabla \cdot \bm{u} = 0,
\end{cases}
\end{equation}
where $\nu$ is the kinematic viscosity, the term $2{\bm \Omega} \times {\bm u}$ is the Coriolis force produced by rotation, and $\bm \Omega = \Omega \hat x_3$ is the angular velocity with frequency $\Omega$ around the rotation axis $\hat x_3$. The fluid density is constant and absorbed into the definition of pressure $p$. The linear friction term $\alpha \Delta^{-1}\bu$, with hypo-viscosity coefficient $\alpha=0.1$ is acting only on wave numbers with $ |\bk|\le 2$ and it is needed to prevent the formation of large-scale a condensate \cite{alexakis2018cascades}.
The forcing mechanism, $\bm{f}$, is a Gaussian process delta-correlated in time, with support in wavenumber space around $k_f=4$. In our DNS we fixed $\Omega=8$, resulting in a Rossby number $Ro = E^{1/2}/k_f \Omega\sim 0.1$, where $E$ is the flow kinetic energy. The dissipation in our simulation is modeled by an hyperviscous term $\nu \nabla^{4} \bm{u}$, which replaces the laplacian in \eqref{eq:nse}, with  $\nu = 1.6\times10^{-6}$. 

\begin{figure}
    \centering
    \includegraphics[width=1.0\textwidth]{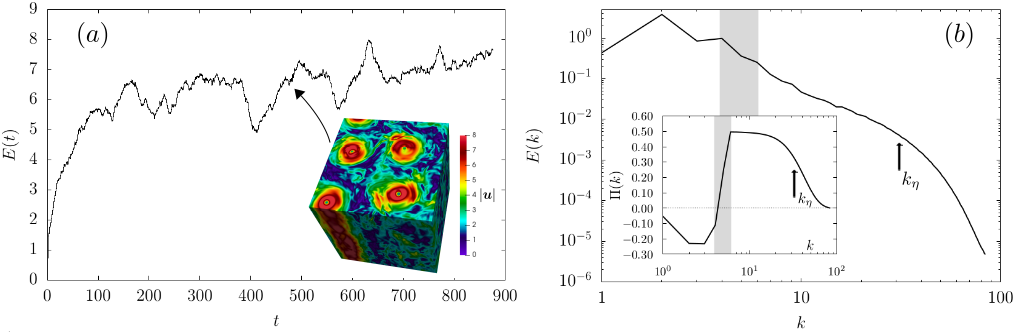}
    \caption{(Panel a) Energy evolution for the turbulent flow generated during the simulation performed to generate the database of $600$ different configurations. The velocity fields are extracted from time $t=276$ up to  $t=876$ every $t = 1$. In the inset we show a 3d visualization of the velocity amplitude $\bu$ extracted from the simulation at a fixed time. (Panel b) Log-log plot of the averaged energy spectrum. Inset: energy flux $\Pi(k)$. The gray areas indicate the forced wavenumbers, while $k_\eta \sim 32$ is the Kolmogorov dissipative wavenumber, defined as the scale such that for  $k >k_\eta$ the velocity field becomes smooth \cite{frisch1995turbulence}.} 
    \label{fig:simulation}
\end{figure}

The total energy evolution as a function of time evolved in the full simulation is shown in Fig.~\ref{fig:simulation}(a), while in the inset of the same figure we show a 3d snapshot of the turbulent velocity field taken at a fixed time in the stationary regime, where we can see large columnar vortices as well as other small-scale structures dispersed in the whole flow domain. 

In Fig.~\ref{fig:simulation}(b) we show the energy spectrum $E(k)= \frac{1}{2} \sum_{ k \le|\bk| < k+1} |\bhu(\bk)|^2$ averaged over time during the simulation, the shaded area  identifies the frequency where the forcing  is acting. It is easy to see how both ends of the spectrum get populated with active modes on  almost two decades in  wave-numbers. This is due to the presence of an energy flux that cascades energy in booth directions, namely towards smaller and higher scales. In the inset of In Fig.~\ref{fig:simulation}(b) we plot the averaged energy flux across the scale $k$, defined as: $ \Pi(k)=-\sum_{|\bk|\le k}  ik_j \hat{u}^*_i(\bk)\sum_{\bp,\bq} \hat{u}_i(\bp)\hat{u}_j(\bq) \delta(\bp+\bq -\bk)$, where the $\delta$ function constrains the nonlinear interactions to closed wave-vectors triads. The negative values of $\Pi(k)$ measured at scales $k<k_f$ indicate that in this range of scales there is an inverse energy cascade, while a positive, $\Pi(k)$ show the presence of a simultaneous forward energy cascade for $k >k_f$ leading to the formation of small-scales structures. 

\section{DataBase Description}
TURB-Rot database is extracted from the simulation described in the previous section as follows:
\begin{itemize}
    \item During the simulation we have dumped a number of $600$ snapshots of the full 3d velocity field (see Fig.~\ref{fig:simulation}(a) for the total energy evolution). Snapshots are chosen with large temporal separation in a way to decrease correlations in time between two successive data-points. {These configurations are the backbone of our image dataset and they can be found under the { data-class $3d\_256$cubed.}}
    \item {We provide also a set of 2d downsized images that can be easily used as input to ML algorithms}, under the name { $2d\_64$squared}. In particular, the original DNS resolution of $256^3$ grid points is downsized to $64^3$ grid, after applying a Galerkin truncation  in Fourier space where the maximum frequency allowed is set to $|\bk| \le  32$. 
    \item For each configuration, $16$ horizontal planes $(x_1,x_2)$ are selected at different $x_3$-levels (the rotation direction).  Each of these $16$ planes is shifted along both $x_1$ and $x_2$ using periodic boundary conditions in $11$ different ways, by choosing randomly a new center of the plane, such as to obtain a total of $16 \times 11 = 176$ planes at each instant of time. 
    \item Finally, the dataset composed of $600 \times 176 = 105600$ planes, is reordered in time randomly to avoid any correlation between successive planes.
\end{itemize}  

\begin{figure}
    \centering
    \includegraphics[width=0.8\textwidth]{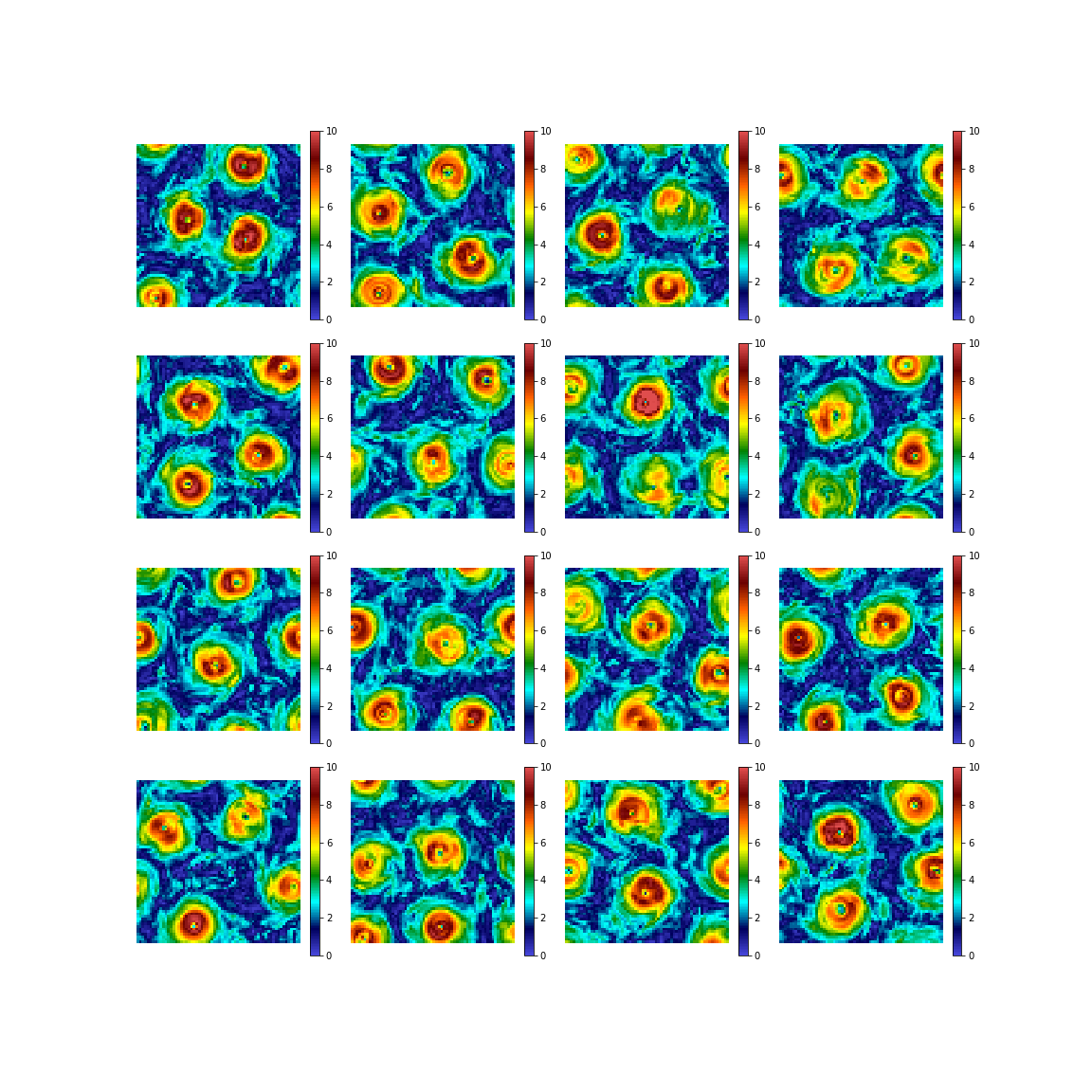}
    \caption{Example of $16$ different $64\times64$ planes composing the TURB-Rot database. The figures presented here are obtained measuring the amplitude of the velocity fields $|\bu|$.}
    \label{fig:database1}
\end{figure}

\begin{figure}
    \centering
    \includegraphics[width=0.8\textwidth]{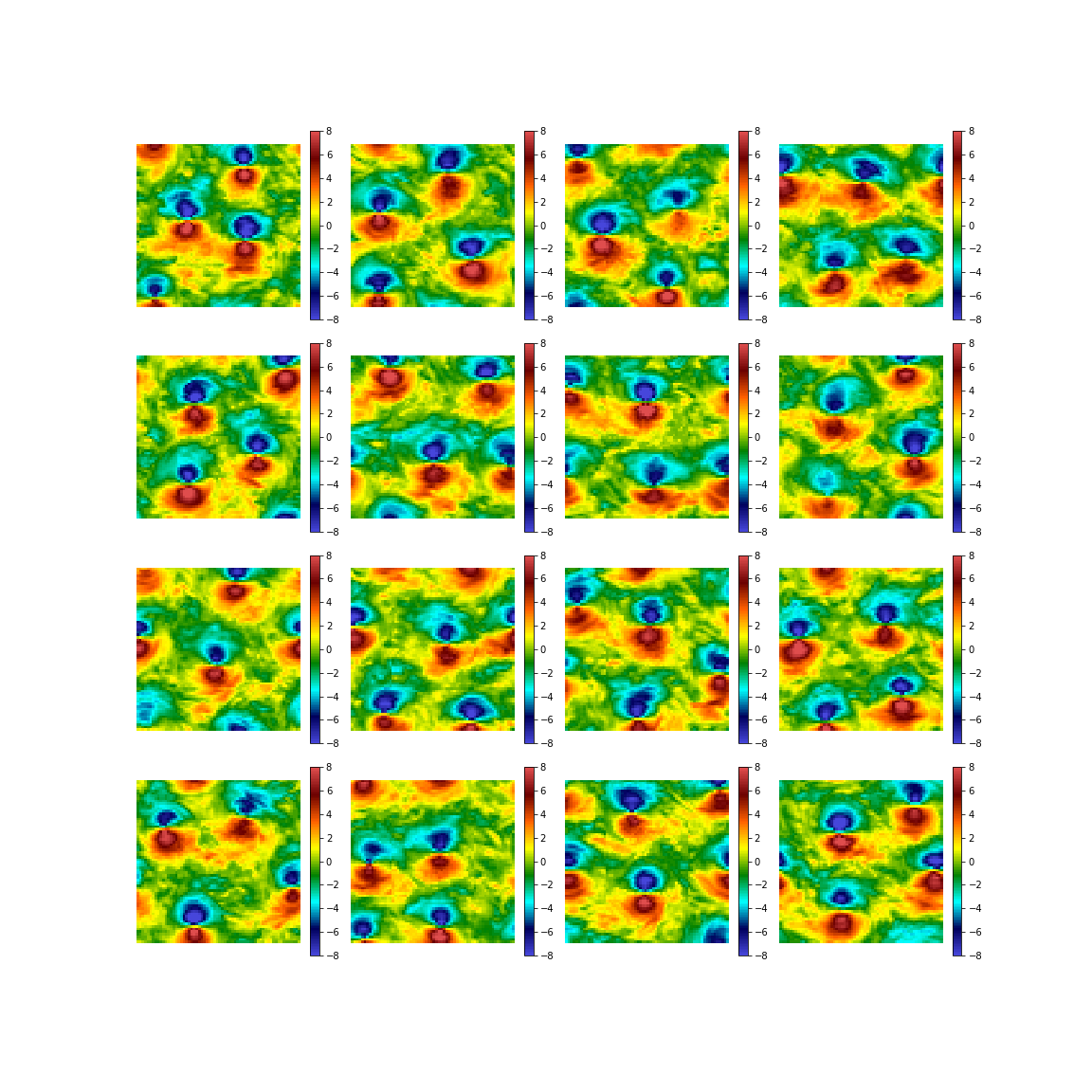}
    \caption{Same planes shown in Fig.~\ref{fig:database1} but for the ${v_1}$ component of the velocity field.}
    \label{fig:database2}
\end{figure}

\begin{figure}
    \centering
    \includegraphics[width=0.8\textwidth]{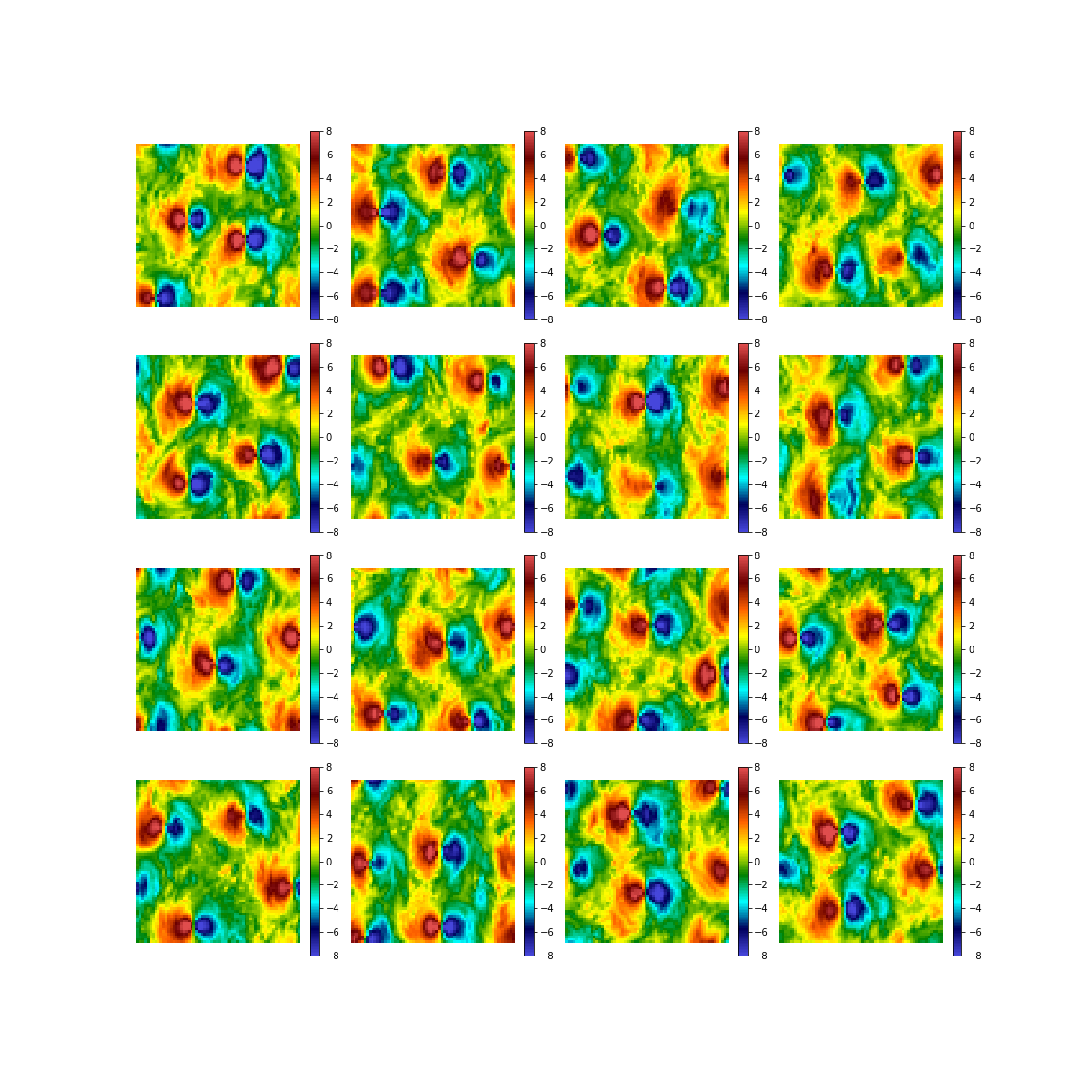}
    \caption{Same planes shown in Fig.~\ref{fig:database1} but for the ${v_2}$ component of the velocity field.}
    \label{fig:database3}
\end{figure}

\begin{figure}
    \centering
    \includegraphics[width=0.8\textwidth]{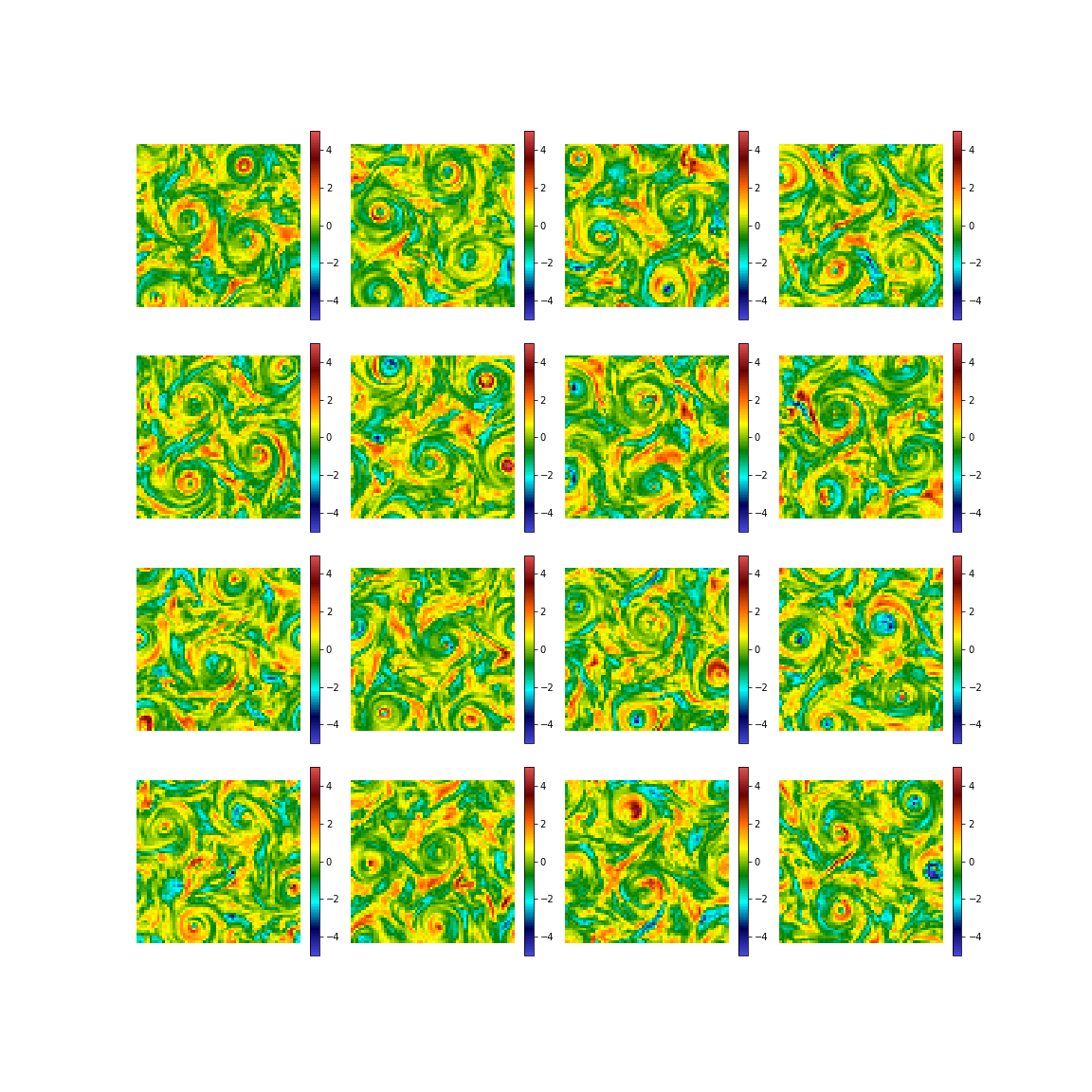}
    \caption{Same planes shown in Fig.~\ref{fig:database1} but for the ${v_3}$ component of the velocity field.} 
    \label{fig:database4}
\end{figure}

\begin{figure}
    \centering
    \includegraphics[width=0.8\textwidth]{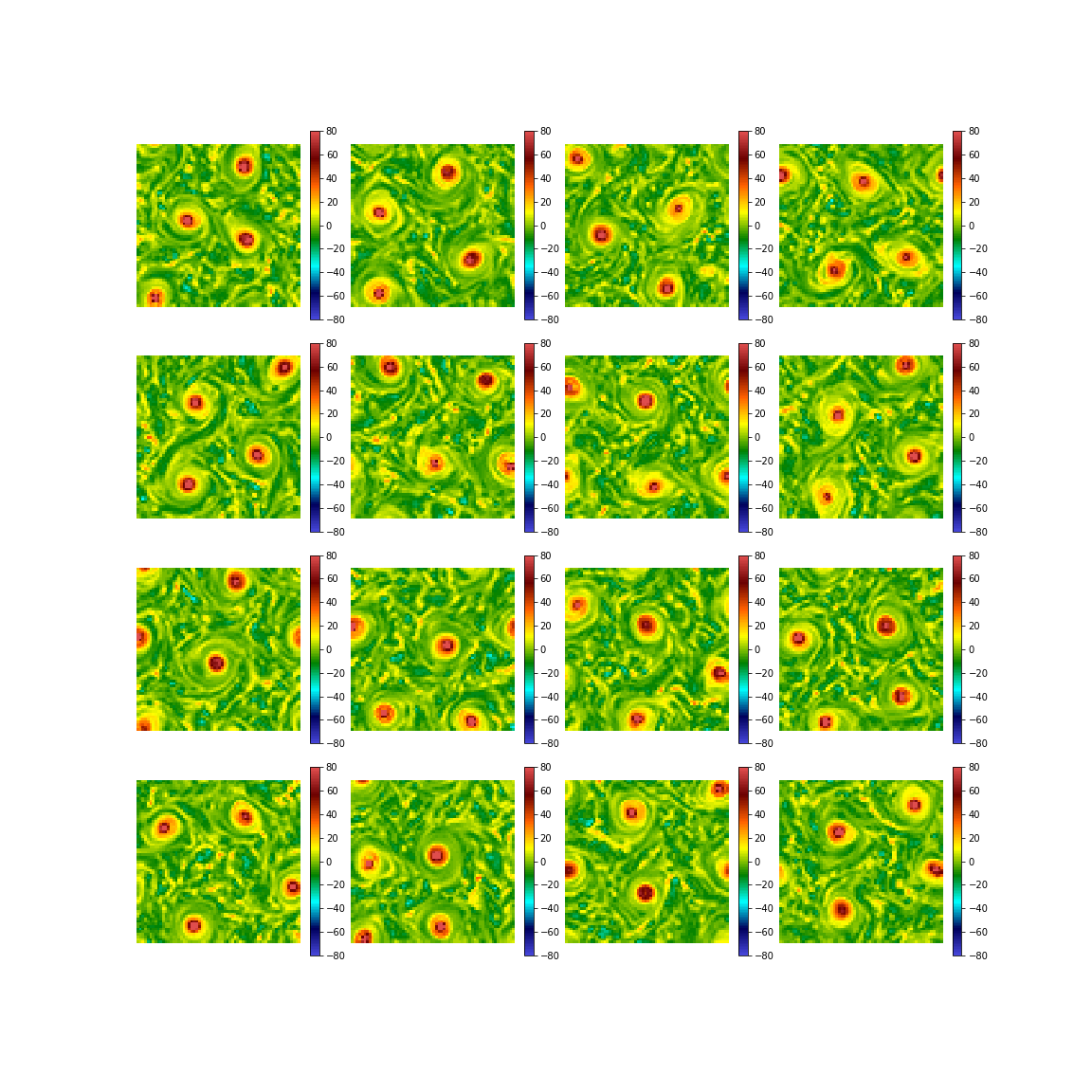}
    \caption{Same planes shown in Fig.~\ref{fig:database1} but for the ${w_3}$ component of the vorticity.} 
    \label{fig:database5}
\end{figure}

The database TURB-Rot is available for download using the SMART-Turb portal \url{http://smart-turb.roma2.infn.it}. The portal is based on  the concept of "Dataset" to aggregate resources related to the same simulation: we have released both the original full resolution of 3d DNS snapshot at $256^3$ grid points and the database of $105600$ ($x_1,x_2$)-planes of size $64 \times 64$ for each velocity component leading to a total of roughly 300K  images. Details on how to access the data with a few examples can be found on the portal. Other data-sets concerning rotating turbulence at higher resolution, different rotation rate, $\Omega$, and including also Lagrangian data for tracers and inertial particles will be made open and available soon.  \\
\noindent Finally, In Fig.~\ref{fig:database1} we show a few examples of the images contained in TURB-Rot made out of the velocity magnitude. In Figs~\ref{fig:database2}-\ref{fig:database4} we show  for the same snapshots the images generated looking at the three different velocity components. In Fig.~\ref{fig:database5} finally, we show the image obtained from the same field but looking at the vorticity field, defined as $w_3 = \partial_1 u_2 - \partial_2 u_1$.


\bibliographystyle{unsrt}
\bibliography{references}  

\end{document}